\documentclass[aps,pra,twocolumn,showpacs,superscriptaddress,groupedaddress,eqsecnum]{revtex4}  
\usepackage{graphicx} 	
\usepackage{dcolumn}  	
\usepackage{bm}              	
\usepackage{amssymb}   	
\usepackage{amsmath}   	
\usepackage{braket}         

\begin{document}
\widetext

\title{Quantum routing games}

\author{Neal Solmeyer}
\affiliation{Sensors and Electron Devices Directorate, Army Research Laboratory, Adelphi, MD, 21005-5069, USA.}
 \author{Ricky Dixon} 
\affiliation{Mississippi Valley State University, 14000 Highway 82 West, Itta Bena, MS, 38941.}

\author{Radhakrishnan Balu }
\affiliation{Computer and Information Sciences Directorate, Army Research Laboratory, Adelphi, MD, 21005-5069, USA. }
\affiliation{University of Maryland Baltimore County, 1000 Hilltop Circle, Baltimore, MD 21250, radbalu1@umbc.edu}

\date{Received: date / Accepted: date}
\begin{abstract}
We discuss the connection between a class of distributed quantum games, with remotely located players, to the counter intuitive Braess' paradox of traffic flow that is an important design consideration in generic networks where the addition of a zero cost edge decreases the efficiency of the network. A quantization scheme applicable to non-atomic routing games is applied to the canonical example of the network used in Braess' Paradox. The quantum players are modeled by simulating repeated game play. The players are allowed to sample their local payoff function and update their strategies based on a selfish routing condition in order to minimize their own cost, leading to the Wardrop equilibrium flow. The equilibrium flow in the classical network has a higher cost than the optimal flow. If the players have access to quantum resources, we find that the cost at equilibrium can be reduced to the optimal cost, resolving the paradox. 
\end{abstract}

\maketitle

It is becoming increasingly important to consider congestion of information in communication networks. In ad-hoc mobile networks, that may change dynamically with nodes that act independently of one another, it can be particularly difficult to efficiently route information. In many contexts, game theory is a powerful tool for analyzing such problems \cite{Han2012}. Game theory is used to solve for the equilibrium flow of information when each node acts independently and routes information selfishly. The equilibrium may not be the desired flow from the point of view of a total global cost. The goal of designing a network protocol is to ensure that the equilibrium flow is close to the optimal flow.

As quantum networking hardware begins to come online \cite{Hucul2015,Nolleke2013}, it may be possible for network games to take advantage of the benefits that quantum games have shown over classical games. By incorporating quantum information into a game setting, quantum games may have different equilibria that can outperform their classical counterpart\cite{Meyer1999}. Entanglement shared between different nodes of the network allows the players to have correlated outcomes even in the absence of communication. This leads to the quantum game sampling a larger space of probability distributions that can realize equilibria resembling classical correlated equilibria which are only possible in classical games when the players receive advice\cite{Auman1974}. 

There have been a few previous examples of quantization schemes of classical routing games. The games have been simplified and the strategy choices restricted so that they map onto the prisoners' dilemma\cite{Zableta2014,Zableta2016} where the well-known quantum solution outperforms the classical one \cite{Eisert1999}. Pigou's example was also mapped onto the prisoners' dilemma where an entangled pair was shared between two players at the same node\cite{Scarpa2010}. Since the entanglement is at one node, this approach cannot not take full advantage of the non-local characteristics of quantum entanglement for larger networks. A notable example used Bell pairs to avoid packet collisions in a software defined networking simulation of an atomic routing game\cite{Hasanpour2017}. Inspired by the deep connection between Bell's inequalities and Bayesian quantum games\cite{Brunner2013}, by mapping the probability distributions of the possible routing paths onto Bell's inequalities, the load-balancing of ad-hoc networks can be improved by using quantum correlations.

The present work aims to provide a more general framework for non-atomic quantum routing games. The Eisert, Wilkens, Lewenstei quantization scheme (EWL)\cite{Eisert1999} is applied to non-atomic routing games and we look at the consequences of selfish routing when there is no restriction on the players' strategy choices. This scheme would be applicable to a classical information network which has an overlaid quantum network. We assume that the nodes of the network share many sets of entangled particles that are distributed prior to the game. Other than the communication required to establish entanglement, the players do not communicate and do not receive advice from the referee.  The players query the quantum network, apply their strategy choices, and make routing decisions for the classical network based on the results of a measurement on the quantum network. We find that the quantum network has an equilibrium under selfish routing that performs better than a purely classical network. This approach resloves the well-known Braess' paradox from classical network theory and serves as a proof of principle that quantum networks may enable more efficient routing on classical networks due to their ability to realize quantum correlations.

Routing games are formulated as a collection of source-sink pairs in a directed graph. Flow on the graph represents traffic or information flow over a communication network. We model a non-atomic game where the information is divided in continuous units, with the total flow from the source to the sink normalized to 1. The players of the game can be interpreted as each infinitesimal unit of information at each node being routed through the network. Since each player acts independently in their own interest, it is natural to analyze this in a game theoretical context. 

Each edge in the graph has a time delay, or latency, $L[f]$, associated with it, which is a function of the flow on that edge, $f$. The latency serves as the cost function of the game. The selfish routing case is where the players try to minimize their own latency, or equivalently, when they try to equalize the latency of each of their outgoing paths \cite{Roughgarden2007}. This leads to the {\sl Wardrop equilibrium} (which is analogous to the { \sl Nash equilibrium } in standard game theory) \cite{Wardrop1952}.  Analytically the equilibrium flow, $\{f_e\}$ is the set of flows that satisfies \cite{Ozdaglar2008}:

\begin{equation}
\text{min}\Big(\sum_j \int_0^{f_j} L_j[z] dz\Big) 
\label{eq:pigou_equilibrium}
\end{equation}
Where the sum is over all $j$ edges or channels.

The total cost of a set of flows can be measured as the average cost for all flows on a network and is given by the sum of the latency on each edge multiplied by the amount of flow on that edge:
\begin{equation}
C_T[f]= \sum_j{f_j L[f_j]} 
\label{eq:total_cost}
\end{equation}

The optimal flow, $\{f_o\}$, is the set of flows that produce the global minimum of Eqn. \ref{eq:total_cost}. This flow is optimal from a societal perspective as it minimizes the average cost. If not equal to the equilibrium flow, the optimal flow is only accessible if the players agree to cooperate through some central mechanism such as shared advice, or a contract. A useful metric in congestion games is the price of anarchy, ($\kappa$) which is the ratio of the total cost at the Wardrop equilibrium to the total cost of the optimum flow $\kappa=C_T[f_e]/C_T[f_o] $. The cost of the optimal flow does not change when the game is quantized, rather, the goal is to structure the game so that the price of anarchy approaches 1.

\begin{figure}
\includegraphics[width=\columnwidth]{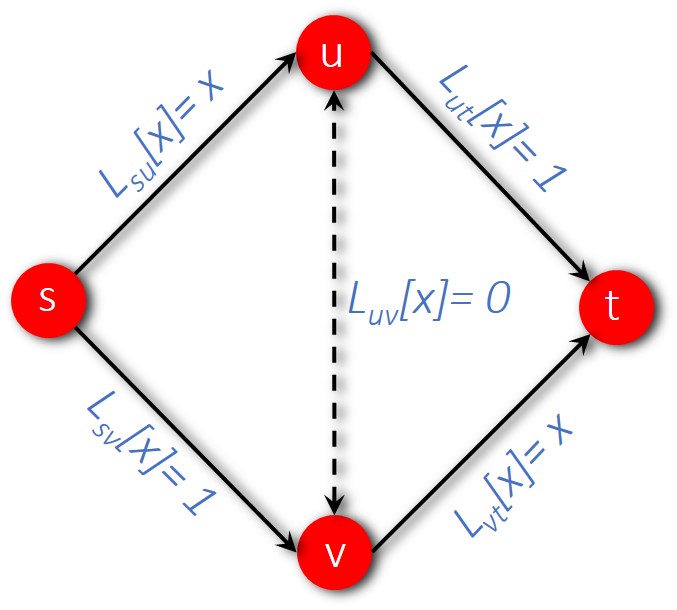}
\caption{\label{fig:Braess} Braess' paradox}
\end{figure}

The counter intuitive Braess' paradox arises from the intuition that a new, zero cost, link in a network will only improve the efficiency of the network.  It is formulated on a four node network with source $s$ and sink $t$, as seen in Fig. \ref{fig:Braess}. The paths $s \rightarrow u$ and $v \rightarrow t$ have a latency equal to the amount of flow on the edge, while the paths $s \rightarrow v$ and $u \rightarrow t$ have constant latency equal to 1. 

If the path through nodes $u$ and $v$ does not exist, the equilibrium flow of the network is with the flow equally shared on the two possible paths, $s \rightarrow u \rightarrow t$ and $s \rightarrow v \rightarrow t$ with a $C_T = 3/2$. This is also the optimal flow, giving the network $\kappa = 1$. If one tries to improve the performance of the network by adding the bi-directional zero cost edge between $u \rightarrow v$ the equilibrium flow actually has a higher total cost of $ C_T = 2$ as each player tries to take advantage of the lowest cost path $s \rightarrow u \rightarrow v \rightarrow t$. The optimal flow is the same as the optimal flow without the central node, and therefore $ \kappa = 4/3$. The paradox is exemplified as the addition of a zero cost node increases the price of anarchy from 1 to 4/3.

Braess' Paradox \cite{Braess1968} is a historically interesting example which has several real world analogues\cite{Roughgarden2006} including communication networks, transport networks \cite{Albert2016}, biology, and even has an analogue in a physical system of strings and springs \cite{Cohen1991}.

To quantize the game in the EWL quantization scheme, one qubit is assigned to each node where a player has to chose which direction to route the information, $s, u$ and $v$ in this case. The state of the players' qubits is initialized to $\ket{000}$. An $N$ qubit entangling operation is performed between the three nodes \cite{Benjamin2001,Du2003}:

\begin{equation}
J_N(\gamma)=e^{i \gamma \sigma_x\otimes^N}
\label{eq:JN}
\end{equation}
The amount of entanglement is parametrized by $\gamma$, which is maximal at $\gamma = \pi/2$ and zero at $\gamma = 0$.

Next, the players at each node apply an arbitrary unitary rotation to their qubits which serves as their strategy choice. An arbitrary rotation can be written:

\begin{equation}
\hat{U}(\theta, \phi,\alpha)=
\begin{pmatrix}
e^{-\imath \phi} Cos(\theta/2) & e ^{\imath \alpha} Sin(\theta/2) \\
- e ^{-\imath \alpha} Sin(\theta/2)&e^{\imath \phi} Cos(\theta/2)
\end{pmatrix}
\label{eq:strat}
\end{equation} 

In practice, we place no restriction on the parameters $\theta, \phi$, or $\alpha$ since a winding in phase only produces redundant strategy choices, which do not compromise our analysis. 

Finally, an un-entangling operation $J^\dagger(\gamma)$ is performed on the qubits and the resulting state is measured. 

For each qubit, the two possible measurement outcomes are associated with the two outgoing paths from the node, and the expectation value of the measured qubit determines the amount of information it routes along each path as a fraction of the information that is incoming to the node. This determines the flow along each of the paths, which is then used to calculate the latencies along each path and the total cost of the flow. Symbolically, the flows are a function of the expectation values of final state, $f[\braket{\psi_f|\psi_f}]$, which is a function of the player's strategy choices and the entangling parameter, $\ket{\psi_f[\theta_s,\phi_s,\alpha_s,\theta_u,\phi_u,\alpha_u,\theta_v,\phi_v,\alpha_v, \gamma]}$. 

To model selfish behavior, we simulate a repeated game where players update their strategy choice based on a 'no-regrets' condition on their local latencies, which should converge on the Wardrop equilibrium \cite{Fischer2006}. The no-regrets condition states that an equilibrium is obtained if no player can improve their payoff by unilaterally altering their strategy choice. Thus, we allow the players to locally sample the cost function as they adjust each of the 3 parameters in their strategy choice in order to approximate the local slope of the cost function to order to update their strategy choice to minimize the difference between latencies of their two outgoing paths. 

After each round, they are given their expected cost function, i.e. for player $s$, the difference of latencies of the outgoing paths is $\delta L_s[\ket{\psi_f}] = L_{su}(f_{su})-L_{sv}(f_{sv})$. Then the local derivative of the cost function is approximated by keeping all other 8 strategy choice parameters fixed, and changing only 1. For example, player $s$ updates the parameter $\theta_s$ by computing $\delta L_{\theta s} = \delta L_s[\ket{\psi_f[\theta_s \rightarrow (\theta_s + d)]}]$, where $d$ is a small parameter. The players then update each of their strategy choices for the $(n+1)^{th}$ round of the game by adjusting the parameter to lower the latency differences of its outgoing paths with a learning parameter, or gain, $M$, i.e.:

\begin{equation}
\theta_s^{(n+1)} =  \theta_s^{(n)} - M(\delta L_s - \delta L_{\theta s} )
\label{eq:update}
\end{equation}

\begin{figure*}
\includegraphics[width=2.0\columnwidth]{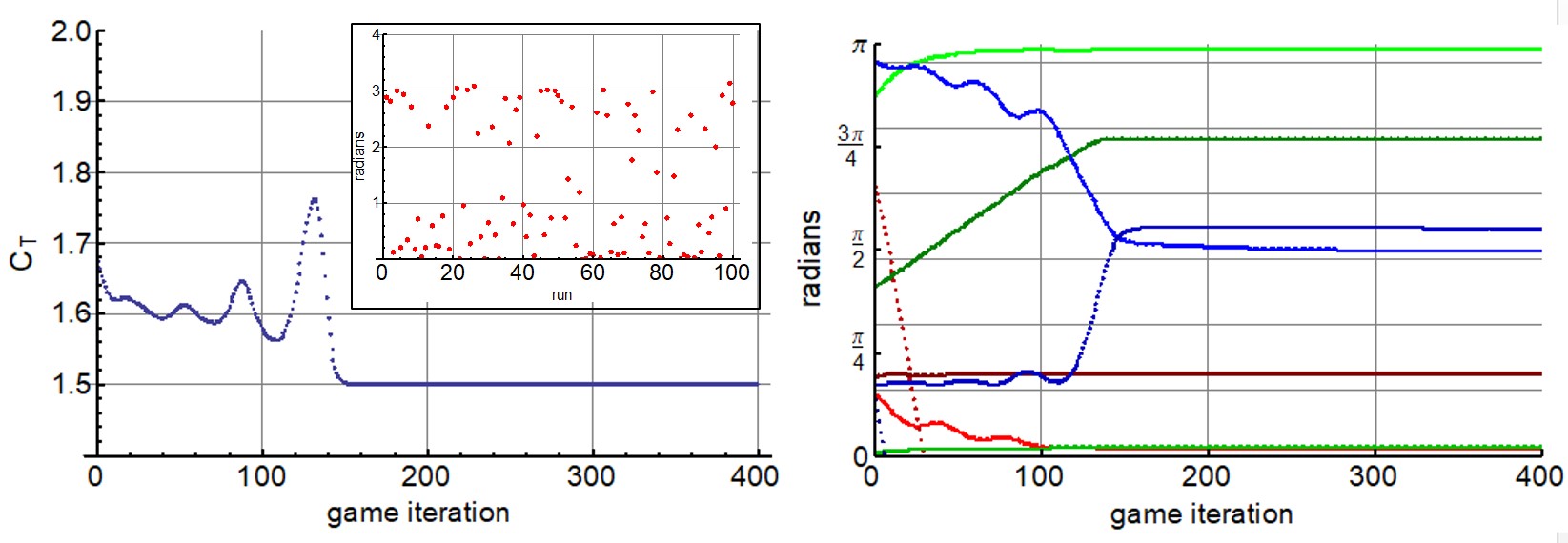}
\caption{\label{fig:instantiation} A typical instantiation of the simulated repeated game for $\gamma = \pi/4$, with a gain of $M = 10$ and $d = 0.01$. Left: the total cost is plotted as a function of the repeated game iteration. It can be seen that total cost converges to an equilibrium, which is close to the optimal flow $C_T = 1.5$. Right:  the value of the various strategy parameters are plotted. Most of the parameters converge to an equilibrium and a few run off. The inset on the left shows the equilibrium value of one of the strategy parameters for repeated runs of the simulation.}
\end{figure*}

This is done for all three strategy parameters $\{\theta,\phi,\alpha\}$ of all three players $\{s,u,v\}$. The players initially choose random  $\{\theta,\phi,\alpha\}$, and the game is repeated until an equilibrium is reached.

We simulate the full network from Fig. \ref{fig:Braess}, including the central 0 cost edge, as the graph with no central edge is already optimal with $\kappa = 1$, and cannot be improved with quantum players. An example run of the simulation is shown on the left side of  Fig. \ref{fig:instantiation}. The simulations are typically is performed with 400 iterations, a gain of $M = 10$, and $d = 0.01$. The example shown is for a partially entangled initial state $\gamma = \pi/4$. The total cost of the flow is plotted on the left, the fact that the total cost comes to a fixed value shows that the algorithm does indeed lead to an equilibrium flow. 

The graph on the right of Fig. \ref{fig:instantiation} plots the values of the 9 strategy parameters. The strategy parameters also come to stable values. Occasionally, strategy parameters can appear to run off and not stabilize, as do two that appear in Fig. \ref{fig:instantiation}. This can be either because they are either irrelevant to the equilibrium, they maintain a fixed difference to another strategy parameter, or they later converge outside of the bounds of the graph. Each of these possibilities were seen in different runs of the simulation. It is possible for parameters to be irrelevant to the game due to the structure of the strategy matrix Eq. \ref{eq:strat}, where for example, when $\theta = 0$, $\alpha$ is undefined.

Each run of the simulation produces different final values for the equilibrium strategy parameters, though the cost at the equilibrium is always the same. The equilibrium value for one of the strategy parameters is shown for 100 different runs in the the inset on the left of Fig. \ref{fig:instantiation}. This is not surprising, as games quantized in the EWL scheme with arbitrary strategy choices have many equilibria, and are often defined as a fixed difference between parameters, rather than the value of the parameters themselves \cite{Solmeyer2017}.

\begin{figure}
\includegraphics[width=\columnwidth]{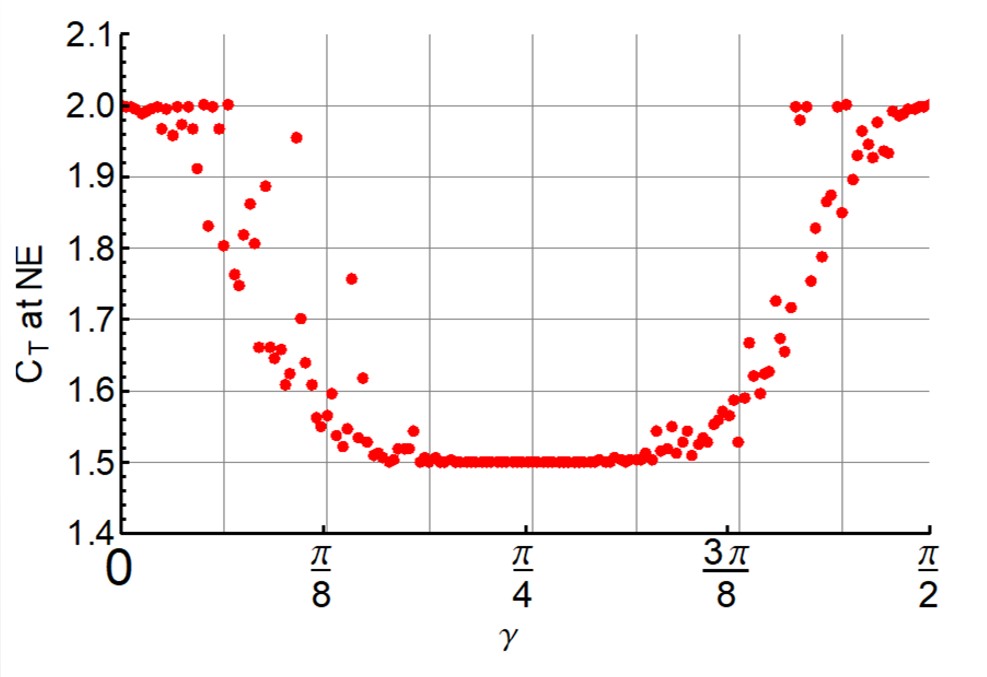}
\caption{\label{fig:quantumPartial} The total cost of the flow after an equilibrium has been reached is plotted as a function of the entanglement, $\gamma$. We simulated 400 iterations of the repeated game, with a gain of $M$ and $d = 0.01$.}
\end{figure}

In Fig. \ref{fig:quantumPartial}, we plot the cost at equilibrium found for simulations with varying amounts of entanglement. For $\gamma = 0$, the equilibrium flow is the same as in the classical case. This is demonstrative of the fact that it is a properly quantized game. This also proves that the Braess' paradox remains in the quantum network even when a much wider set of strategy choices is allowed. The total cost at equilibrium has a minimum at $\gamma = \pi/4$ and is equal to the optimal cost, which resolves the paradox. It is interesting to note that at maximal entanglement, Braess' paradox is recovered, as the equilibrium flow again goes to the value in the classical game, with a price of anarchy approaching 1.33. The optimal value of entanglement to take full advantage of the quantum correlations is not the maximal entanglement, but rather, half entanglement. 

At $\gamma = \pi/4$, though the strategy choices may be different for each run of the simulation, the flow always converges to 0.5 flow on all edges (except for $u \rightarrow v$ which has $f_{uv} = f_{vu} = 0 $), and thus has a total cost at equilibrium equal to the optimal cost and $\kappa = 1$. 

For comparison, other networks were simulated with the same topology but different edge latencies. We sampled constant, linear, and quadratic latencies for the various edges. If, for a given network, the price of anarchy in the classical game was unity, $\kappa_C = 1$, it was in the quantum game as well, i.e. $\kappa_Q = 1$. In cases with a classical price of anarchy $\kappa_C>1$, the quantum version had a price of anarchy $ \kappa_C> \kappa_Q > 1$. For the networks which are symmetric such that $L_{su} = L_{sv}$ and $L_{ut} = L_{vt}$, the quantum equilibrium is optimal, i.e. $\kappa_Q = 1$. When the network is asymmetric, the optimal flow does not have equal flows on all channels. Yet, the new quantum equilibrium flow stall has $f = 0.5$ in all channels except the central channel. As a result, the cost at equilibrium is not optimal, though it does outperform the classical one.  These simulations show that the quantum network performs at least as well as the classical network, and many cases better. Further, even in the presence of asymmetry, the new symmetric quantum equilibrium flow is still stable and performs better than the classical equilibrium.

The present scheme relies on the distribution of entangled particles between the nodes of the network and a central referee. This certainly requires communication, though the type of communication required is only to establish the entanglement, and does not broadcast any information about the players' intentions. After this entanglement is established, there is no communication necessary to establish the quantum correlations between the remote nodes.

This scheme does require the nodes to transmit their quantum particle to the referee in order for the un-entangling operation, though by interlacing additional qubits in a manner similar to quantum cryptography schemes, it should be possible to incorporate the ideas of physical security enabled by quantum information principles into these protocols. So that, even though some form of connection between nodes must remain in place, that connection can be made to be secure against attacks. The security aspects of quantum information could provide additional motivations for incorporating quantum game ideas into classical networking protocols.

Though something resembling the present scheme would likely require far more quantum resources than will be available in the near future, it serves as a proof of principle that quantum game ideas may be useful in improving the efficiency of classical networks when the players are allowed to act in their self interest. Another important consideration that work demonstrates is that a quantum network can exhibit the Braess' Paradox even for maximally entangled states. Thus, quantum networking schemes may necessarily have to incorporate these types of counter intuitive behaviours into their analysis of network performance. When quantum networks come online, quantum game theory may provide novel solutions to common networking problems and may become a necessary tool for analyzing the behavior of the network.


\begin{thebibliography}{99}

\bibitem
{Han2012} Z. Han, D. Niyato, W. Saad, T. Baar, and A. Hjrungnes, "Game theory in Wireless Communication Networks: Theory, Models, and Applications", Cambridge University Press, New York, (2012).

\bibitem{Hucul2015}
D. Hucul, {\sl et al.}, Nature Phys. {\bf 11}, pp. 37-42 (2015)

\bibitem{Nolleke2013}
C. N\"{o}lleke, {\sl et al.}, Phys. Rev. Lett. {\bf 110}, 140403 (2013)

\bibitem
{Meyer1999}Meyer, D.:, Phys. Rev. Lett. 82, 1052-1055 (1999). 


\bibitem{Auman1974} 
Auman, R., Journal of Mathematical Economics, 1, p. 67-96, (1974).


\bibitem{Zableta2014}
Zableta, O. G., M. Arizmendi, Journal of advances in applied and computational mathematics, Vol. 1, No. 1, 2014.

\bibitem{Zableta2016}
Zableta, O. G., J. P. Barrangu, and C. M. Arizmendi, ArXiv preprint: 1608.07264v1

\bibitem{Eisert1999}
Eisert, J., M. Wilkens and Lewenstein, M., Phys. Rev. Lett. 83, 3077-3080 (1999) 

\bibitem{Scarpa2010}
Scarpa, G., Nerwork games with quantum strategies, QuantumCom 2009, LNICST 36, Eds: A. Sergienko, S. Pascazio and P. Villoresi, pp 74-81. 


\bibitem{Hasanpour2017}
Hasanpour, M., {\sl et al.}, Quantum inf. Process. 16, 148 (2017).

\bibitem{Brunner2013}
Brunner, N. and Linden, N., Nature Communications, {\bf 4}, 2057 (2013). 

\bibitem{Roughgarden2007} 
T. Roughgarden, Routing Games, in "Algorithmic game theory," Ed. N. Nisan, T. Roughgarden, E. Tardos, and V. V. Vazirani, Cambridge University Press, Cambridge, pp. 461-484, (2007).


\bibitem{Wardrop1952}
J. G. Wardrop, Some Theoretical aspects of road traffic research, in {\sl Proc. Institute of Civil Engineers, pt. II,} vol 1, pp. 325-378, (1952).

\bibitem{Ozdaglar2008}
A. Ozdaglar, Networks' challenge: where game theory meets network optimization, International symposium on information theory, 2008.

\bibitem{Braess1968}
D. Braess, {\"U}ber ein Paradoxon aus der Verkehrsplanung. {\sl Unternehmensforschung,} 12:258-268, (1968); D. Braess, Transport. Sci., 39(4)446-450, (2005).
 
\bibitem{Roughgarden2006}
T. Roughgarden, J. Computer System Sci. 72(5): 922-953, (2006).

\bibitem{Albert2016} Albert Sol{\'e}-Ribalta, Sergio G{\'o}mez, and Alex Arenas, Phys. Rev. Lett. 116, 108701(2016).

\bibitem{Cohen1991}
J. Cohen and P. Horowitz,  Nature, 352 (8): 699-701, (1991).

\bibitem{Benjamin2001}
Benjamin, S. C., and P. M. Hayden, Phys. Rev. A, 64, 030301(R) (2001). 

\bibitem{Du2003}
Du, J. {\sl et al.}, Phys. Letters A, 302, p. 229-233 (2002).

\bibitem{Fischer2006}
S. Fischer, H. R{\"a}cke, and B. V{\"o}cking, Fast Convergence to Wardrop equilibria by adaptive sampling methods. in {\sl Proc 38th Symp. Theory of Computing,} pp. 653-662, (2006).

\bibitem{Solmeyer2017} N.\ Solmeyer, R.\ Dixon, and R.\ Balu, Quantum Inf. Process. \textbf{16}, 146 (2017).



\end{thebibliography}
\end{document}